\documentclass[10pt]{iopart}

\usepackage{iopams}
\usepackage{amssymb}
\usepackage{amsfonts}
\usepackage{amstext}
\usepackage{graphicx}
\usepackage{setstack}
\usepackage{amscd}
\usepackage{color}

\newcommand{\ihbar}{\imath \hbar}
\newcommand{\Lin}{\mathrm{Lin}}

\renewcommand{\Im}{\mathrm{Im}}

\begin{document}

\title[Geometric phases for nonselfadjoint hamiltonians]{Consistency between adiabatic and nonadiabatic geometric phases for nonselfadjoint hamiltonians}

\author{David Viennot, Arnaud Leclerc \& Georges Jolicard}
\address{Institut UTINAM (CNRS UMR 6213, Universit\'e de Franche-Comt\'e), 41bis Avenue de l'Observatoire, BP1615, 25010 Besan\c con cedex, France.}
\author{John P. Killingbeck}
\address{Centre for Mathematics, University of Hull, Hull HU6 7RX, UK.}

\begin{abstract}
We show that the adiabatic approximation for nonselfadjoint hamiltonians seems to induce two non-equal expressions for the geometric phase. The first one is related to the spectral projector involved in the adiabatic theorem, the other one is the adiabatic limit of the nonadiabatic geometric phase. This apparent inconsistency is resolved by observing that the difference between the two expressions is compensated by a small deviation in the dynamical phases.
\end{abstract}

\section{Introduction}
The adiabatic approximation is currently a commonly used tool to study dynamical systems \cite{messiah}. Recently Marzlin and Sanders have pointed out a possible inconsistency in the application of the adiabatic theorem \cite{marzlin}. Their work has produced some debate and controversy \cite{wu,sarandy,pati,amin} about the application of the adiabatic approximation to systems governed by selfadjoint Hamiltonians. Generalizations of the adiabatic theorem have also been proposed for nonselfadjoint Hamiltonians by Nenciu and Rasche \cite{nenciu}, Abou Salem and Fr\"ohlich \cite{abou,abou2}, Joye \cite{joye} and Avron {\it et al} \cite{avron,avron2}. In the present work we show that another kind of apparent inconsistency arises for nonselfadjoint Hamiltonians. Like the Marzlin-Sanders inconsistency it is intimately associated with the geometric phase concept. The geometric phase commonly used in the adiabatic approximation (the so-called Berry phase) does not coincide with the adiabatic limit of the geometric phase commonly used in the nonadiabatic cyclic quantum dynamics (the so-called Aharonov-Anandan phase) \cite{aharonov,page}.  Section 2 presents general considerations about the two possible expressions for the geometric phase. Section 3 is an analysis of the origin of the apparent inconsistency. We show how to treat correctly the adiabatic approximation of the nonadiabatic geometric phase to resolve this inconsistency.

\section{General considerations}
\begin{figure}
\begin{center}
\includegraphics[width=9cm]{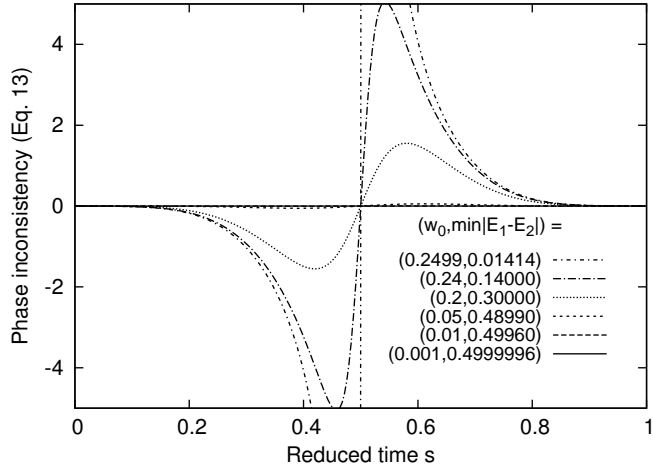}
\caption{\label{fig1} $\langle \phi^*_1|\dot \phi_1 \rangle - \frac{\langle \phi_1|\dot \phi_1 \rangle}{\langle \phi_1|\phi_1\rangle}$ for the hamiltonian $H(s) = \left( \begin{array}{cc} 0 & \Omega(s) \\ \Omega(s) & - \imath \frac{\Gamma}{2} \end{array} \right)$ with $\Omega(s) = \Omega_0 e^{-\frac{(s-s_0)^2}{2 \sigma}}$, with different values of $w_0 = \frac{\Omega_0}{\Gamma}$. The gaussian parameters are $s=0.5$ and $\sigma = 0.16$. $\min|E_1-E_2|$ is the minimal distance during the evolution between the two eigenvalues (this distance is proportionnal to the inverse of nonadiabatic couplings). This model is associated with a quantum bound state coupled to a quantum resonance (with resonance width $\Gamma$) by a laser gaussian pulse.}
\end{center}
\end{figure}

Let $H(s)$ be a $\mathcal C^1$-time-dependent nonselfadjoint hamiltonian with $\frac{1}{2\imath}(H(s)-H(s)^\dagger) \leq 0$ ($H(s)$ generates a contraction). $s=\frac{t}{T}$ is the reduced time, $T$ is the total duration. Let $\lambda_a(s) \in \mathbb C$ be an isolated non-degenerate eigenvalue of $H$, $\phi_a(s)$ be the associated (right) $\mathcal C^1$-eigenvector and $\phi_a^*(s)$ be the associated biorthogonal (left) $\mathcal C^1$-eigenvector:
\begin{eqnarray}
H \phi_a & = & \lambda_a \phi_a \\
H^\dagger \phi_a^* & = & \overline{\lambda_a} \phi_a^*
\end{eqnarray}
(here the ordinary complex conjugate is denoted by an overline rather than by a star) with
\begin{equation}
\langle \phi_a^*|\phi_b \rangle = \delta_{ab}
\end{equation}
The adiabatic approximation states that the wave function $\psi(s)$, which is the solution of the Schr\"odinger equation $\frac{\ihbar}{T} \dot \psi = H \psi$ with $\psi(0)=\phi_a(0)$, remains approximately projected onto $\Lin(\phi_a)$ ($\Lin$ denotes the linear envelope). $s =\frac{t}{T}$ is the reduced time, $T$ being the total duration. The dot denotes the derivative with respect to $s$. We should first point out that there are two ``natural'' projections onto $\Lin(\phi_a)$, the orthogonal projection:
\begin{equation}
P_o = \frac{|\phi_a \rangle \langle \phi_a|}{\langle \phi_a|\phi_a\rangle}
\end{equation}
and the spectral (Riesz) projection:
\begin{equation}
P_s = \frac{1}{2 \pi \imath} \oint_{\Gamma_{\lambda_a}} (H-z)^{-1}dz = |\phi_a \rangle \langle \phi_a^*|
\end{equation}
where $\Gamma_{\lambda_a}$ is a closed path in the complex plan surrounding only $\lambda_a$. We note that the two projectors satisfy $P_o^2 = P_o$, $P_s^2 = P_s$, $P_sP_o = P_o$ and $P_oP_s = P_s$ but $P_o^\dagger= P_o$ whereas $P_s^\dagger \not= P_s$. The adiabatic theorems of Nenciu-Rasche \cite{nenciu}, Abou Salem-Fr\"ohlich \cite{abou} and Joye \cite{joye} deal with the spectral projector:
\begin{equation}
\label{adiab}
U_T(s,0) P_s(0) = P_s(s) U_T(s,0) + \mathcal O(\frac{1}{T})
\end{equation}
where $U_T(s,0)$ is the evolution operator ($\frac{\ihbar}{T} \dot U_T(s,0) = H(s)U_T(s,0)$ with $U(0,0) = 1$). Equation \ref{adiab} constitutes the fundamental assumption of this work. By construction we then have
\begin{eqnarray}
\psi(s) & = & U_T(s,0) \phi_a(0) \\
& = & P_s(s) U_T(s,0) \phi_a(0) + \mathcal O(\frac{1}{T}) \\
& = & \underbrace{\langle \phi_a^*(s)|U_T(s,0)|\phi_a(0)}_{c(s)} \rangle \phi_a(s) + \mathcal O(\frac{1}{T})
\end{eqnarray}
where $c(s) \in \mathbb C$ is a time-dependent complex coefficient (in contrast with the selfadjoint case, $c$ is not just a phase, since the evolution is not unitary). By inserting the expression $\psi \simeq c \phi_a$ in the Schr\"odinger equation, we find that
\begin{equation}
\label{eqc}
\dot c \phi_a \simeq - (\ihbar^{-1} T \lambda_a \phi_a + \dot \phi_a) c
\end{equation}
By projection of eq. \ref{eqc} with $\langle \phi_a^*|$ we find that
\begin{eqnarray}
\label{psis}
\psi(s) & \simeq & e^{-\ihbar^{-1}T\int_0^s \lambda_ads - \int_0^s \langle \phi_a^* | \dot \phi_a \rangle ds} \phi_a(s) \\
& & \qquad \equiv \psi_s(s) \nonumber
\end{eqnarray}
This is the expression that we can find in the literature concerning the adiabatic geometric phases of nonselfadjoint hamiltonians \cite{montragon,mostafazadeh,mailybaev,mehri,viennot}. Since the norms are not fixed to 1, the gauge structure associated with the geometric phases in the nonselfadjoint case deals with changes of both phase and norm (``geometric factor'' in place of ``geometric phase'' would be a more appropriate expression). In other words, the pincipal bundle describing the geometric phases has not $U(1)$ as structure group but $\mathbb C^*$ (the group of non-zero complex numbers). The expression (\ref{psis}) seems to be quite natural, since the adiabatic theorems deal with the spectral projection. Nevertheless, nothing forbids the projection of eq. \ref{eqc} with $\frac{\langle \phi_a|}{\langle \phi_a|\phi_a \rangle}$, and in this case we find that
\begin{eqnarray}
\label{deviation}
\psi(s) & \simeq & e^{-\ihbar^{-1}T \int_0^s \lambda_a ds - \int_0^s \frac{\langle \phi_a | \dot \phi_a \rangle}{\langle \phi_a | \phi_a \rangle} ds} \phi_a(s) \\
& & \qquad \equiv \psi_o(s) \nonumber
\end{eqnarray}
The apparent inconsistency arises from the adiabatic geometric phases:
\begin{eqnarray}
\langle \phi_a^* | \dot \phi_a \rangle - \frac{\langle \phi_a | \dot \phi_a \rangle}{\langle \phi_a|\phi_a \rangle} & = & \langle \phi_a^* | \dot P_o | \phi_a \rangle \\
& = & -\frac{\langle \phi_a | \dot P_s | \phi_a \rangle}{\langle \phi_a | \phi_a \rangle} \\
& \not= & 0
\end{eqnarray}
This problem does not arise for selfadjoint hamiltonians where $\phi_a = \phi_a^*$. This deviation is moreover proportional to the amplitude of the instantaneous non-adiabatic couplings (see fig. \ref{fig1}). 
It is thus small where the nonadiabatic couplings are small, i.e. far from the eigenvalue crossings. The question is then: What is the correct adiabatic geometric phase to use for nonselfadjoint hamiltonians? An evident argument in favour of the ``spectral adiabatic geometric phase'' is that it is the only one which is compatible with a late application of the adiabatic approximation. Indeed, let $\psi(s) = \sum_b c_b(s) \phi_b(s)$ (we suppose for the sake of simplicity that $H(s)$ is diagonalizable). By putting this expression in the Schr\"odinger equation and by projecting with $\langle \phi_a^*|$ we find
\begin{equation}
\dot c_a = - \ihbar^{-1}T \lambda_a c_a - \sum_b \langle \phi_a^*|\dot \phi_b \rangle c_b
\end{equation}
By then applying the following adiabatic approximation (for $b \not= a$):
\begin{eqnarray}
\langle \phi_a^*| \dot \phi_b \rangle & = & \frac{\langle \phi_a^* | \dot H | \phi_b \rangle}{\lambda_b - \lambda_a} \\
& \simeq & 0
\label{approx}
\end{eqnarray}
we find again that $\psi(s) \simeq \psi_s(s)$. In contrast, by projecting with $\frac{\langle \phi_a|}{\langle \phi_a|\phi_a \rangle}$, since the eigenvectors are not orthogonal, we find
\begin{equation}
\dot c_a = -\ihbar^{-1}T \sum_b \lambda_b c_b \frac{\langle \phi_a |\phi_b \rangle}{\langle \phi_a|\phi_a \rangle} - \sum_b c_b  \frac{\langle \phi_a |\dot \phi_b \rangle}{\langle \phi_a|\phi_a \rangle} - \sum_{b\not=a} \dot c_b \frac{\langle \phi_a |\phi_b \rangle}{\langle \phi_a|\phi_a \rangle}
\end{equation}
An adiabatic approximation seems to be inappropriate to treat this expression and cannot be used to claim that $\psi(s) \simeq \psi_o(s)$. Is this argument sufficient to claim that the ``orthogonal adiabatic geometric phase'' is irrelevant? It seems that the answer is ``no''. First, the rigorously proved adiabatic theorems concern the approximation of eq. \ref{adiab} and not that of eq. \ref{approx} (moreover the use of the approximation eq. \ref{approx} is not efficient even for some selfadjoint cases, see \cite{amin}, and we can remark that the conditions of validity of eq. \ref{approx} are the same as requiring the deviation between geometric phases to be small). More importantly, the orthogonal adiabatic geometric phase is the adiabatic limit of the nonadiabatic geometric phase. Indeed consider a quantum dynamics $\frac{\ihbar}{T} \dot \psi = H(s) \psi(s)$ such that $\psi(1) = \mu \psi(0)$ with $\mu \in \mathbb C^*$ (the dynamics is said to be cyclic). Let $\underline \psi_T(s) \in \Lin(\psi(s))$ be such that $\underline \psi_T(1) = \underline \psi_T(0) = \psi(0)$ ($\underline \psi_T$ is an arbitrary choice in $\Lin(\psi(s))$, called a local section in the geometric language of the fibre bundle theory). By construction, there exists $f(s) \in \mathbb C^*$ such that $\underline \psi_T(s) = f(s) \psi(s)$. By inserting $\psi(s) = f(s)^{-1} \underline \psi_T(s)$ in the Schr\"odinger equation, we find
\begin{equation}
\label{geocycl}
f^{-1} \dot f \underline \psi_T = \ihbar^{-1} T H \underline \psi_T + \dot{\underline \psi}_T
\end{equation}
By projecting this equation on $\underline \psi_T$ we find
\begin{equation}
f^{-1} \dot f = \ihbar^{-1}T \frac{\langle \underline \psi_T |H|\underline \psi_T \rangle}{\langle \underline \psi_T|\underline \psi_T \rangle} + \frac{\langle \underline \psi_T | \dot{\underline \psi}_T \rangle}{\langle \underline \psi_T|\underline \psi_T \rangle}
\end{equation}
Finally
\begin{equation} 
\psi(s) = e^{- \ihbar^{-1}T \int_0^s \frac{\langle \underline \psi_T |H|\underline \psi_T \rangle}{\langle \underline \psi_T|\underline \psi_T \rangle} ds - \int_0^s \frac{\langle \underline \psi_T | \dot{\underline \psi}_T \rangle}{\langle \underline \psi_T|\underline \psi_T \rangle} ds} \underline \psi_T(s)
\end{equation}
We note that no approximation occurs in this last expression. $\frac{\langle \underline \psi_T | \dot{\underline \psi}_T \rangle}{\langle \underline \psi_T|\underline \psi_T \rangle}$ generates the nonadiabatic geometric phase. The nonadiabatic geometric phase has an important property concerning the non-unitary evolution. Since
\begin{equation}
\overline{\frac{\langle \underline \psi_T|\dot{\underline \psi}_T \rangle}{\langle \underline \psi_T|\underline \psi_T \rangle}} = - \frac{\langle \underline \psi_T|\dot{\underline \psi}_T \rangle}{\langle \underline \psi_T|\underline \psi_T \rangle} + \frac{d}{ds} \ln \langle \underline \psi_T | \underline \psi_T \rangle
\end{equation}
we find that
\begin{eqnarray}
\left| e^{- \int_0^s \frac{\langle \underline \psi_T | \dot{\underline \psi}_T \rangle}{\langle \underline \psi_T|\underline \psi_T \rangle} ds} \right|^2 & = & e^{- \int_0^s \frac{d}{ds} \ln \langle \underline \psi_T|\underline \psi_T \rangle} \\
& = & \frac{\langle \psi(0)|\psi(0) \rangle}{\langle \underline \psi_T(s)|\underline \psi_T(s) \rangle}
\end{eqnarray}
and then
\begin{equation}
\|\psi(s)\|^2 = \|\psi(0)\|^2 e^{2 \hbar^{-1}T \int_0^s \Im \frac{\langle \underline \psi_T |H|\underline \psi_T \rangle}{\langle \underline \psi_T|\underline \psi_T \rangle} ds}
\end{equation}
The evolution of the norm (and thus the dissipation evolution) depends only on the dynamical phase. At the end of the evolution, the nonadiabatic geometric phase does not take part in the dissipation process. In this sense it is a good generalization of a ``phase'' for the nonselfadjoint dynamics. During the evolution it belongs to $\mathbb C^*$ and not $U(1)$ (it is not a pure phase), but it corresponds to a right ``anholonomy'' for the cyclicity of the dynamics independently of the dissipation.\\
At the adiabatic limit $T \to + \infty$, it is clear that we can choose the local section such that $\lim_{T \to + \infty} \underline \psi_T(s) = \phi_a(s)$. We have then
\begin{equation}
\lim_{T \to + \infty} \frac{\langle \underline \psi_T | \dot{\underline \psi}_T \rangle}{\langle \underline \psi_T|\underline \psi_T \rangle} = \frac{\langle \phi_a|\dot \phi_a \rangle}{\langle \phi_a | \phi_a \rangle}
\end{equation}
The orthogonal adiabatic geometric phase has the same property as the nonadiabatic geometric phase: it does not take part in the dissipation process and is thus a good generalization of a ``phase'' for the nonselfadjoint adiabatic dynamics, in contrast with the spectral geometric phase, for which we have
\begin{equation}
\|\psi(s)\|^2 \simeq \|\phi_a(0)\|^2 e^{2 \hbar^{-1} T \int_0^s \Im \lambda_a ds} \left|e^{- \int_0^s \langle \phi_a^*|\dot \phi_a \rangle ds}\right|^2
\end{equation}
where
\begin{equation}
\left|e^{- \int_0^s \langle \phi_a^*|\dot \phi_a \rangle ds}\right|^2 = \left|e^{-\int_0^s \langle \phi_a^*|\dot P_o|\phi_a \rangle ds} \right|^2 \not= 1
\end{equation}
The spectral adiabatic geometric phase includes a geometric contribution to the dissipation, which is precisely its deviation from the orthogonal adiabatic geometric phase.

\section{Consistency between the two adiabatic geometric phases}
To solve the apparent inconsistency, we remark first that equation \ref{geocycl} can be projected onto all $\chi(s)$ such that $\langle \chi|\underline \psi_T \rangle \not=0$:
\begin{equation}
f^{-1}\dot f = \ihbar^{-1}T \frac{\langle \chi|H|\underline \psi_T \rangle}{\langle \chi|\underline \psi_T \rangle} + \frac{\langle \chi|\dot{\underline \psi}_T \rangle}{\langle \chi|\underline \psi_T \rangle}
\end{equation}
This induces no inconsistency since by construction for any $\chi$ nonorthogonal to the dynamics we have
\begin{equation}
\ihbar^{-1}T \frac{\langle \chi|H|\underline \psi_T \rangle}{\langle \chi|\underline \psi_T \rangle} + \frac{\langle \chi|\dot{\underline \psi}_T \rangle}{\langle \chi|\underline \psi_T \rangle} =  \ihbar^{-1}T \frac{\langle \underline \psi_T |H|\underline \psi_T \rangle}{\langle \underline \psi_T|\underline \psi_T \rangle} + \frac{\langle \underline \psi_T | \dot{\underline \psi}_T \rangle}{\langle \underline \psi_T|\underline \psi_T \rangle}
\end{equation}
The modification of the geometric phase is compensated by a modification of the dynamical phase (so that the sum of the geometric and dynamical phases is invariant). We note that $\frac{\langle \chi|\dot{\underline \psi}_T \rangle}{\langle \chi|\underline \psi_T \rangle}$ lacks the good property of nonparticipation in the dissipation and has no clear physical sense. Nevertheless we can choose $\chi(s) = \phi_a^*(s)$ (for $T$ sufficiently large, by the adiabatic assumption, $\phi_a^*$ is not orthogonal to the dynamics), and have a geometric phase tending to the spectral adiabatic geometric phase. The inconsistency arises from the fact that all generators of dynamical phases tend to $\lambda_a(s)$.\\
To solve this problem it is important to note that the adiabatic theorems for nonselfadjoint hamiltonians need (like the Joye theorem or the Nenciu-Rasche theorem) a ``superadiabatic renormalization'' \cite{joye,nenciu}. In other words, these theorems do not deal with $\phi_a$ but with $\phi_{aT}^{(1)}$, which is eigenvector of the superadiabatic renormalized Hamiltonian:
\begin{equation}
H_T^{(1)}(s) = H(s) - \frac{\ihbar}{T} \left(\dot P_s P_s + \sum_{b \not=a} \dot Q_{sb} Q_{sb} \right)
\end{equation}
where $\{Q_{sb}\}$ are the spectral projectors onto the other eigenspaces. We note that the demonstrations of the adiabatic theorems for nonselfadjoint hamiltonians need iterations of the superadiabatic renormalization ($H_T^{(n)} = H_T^{(n-1)} - \frac{\ihbar}{T} \left(\dot P_{sT}^{(n-1)} P_{sT}^{(n-1)} + \sum_{b \not=a} \dot Q_{sbT}^{(n-1)} Q_{sbT}^{(n-1)} \right)$). For the present analysis the first step is sufficient. The adiabatic approximation is then
\begin{equation}
\underline \psi_T(s) \sim \phi_{aT}^{(1)}
\end{equation}
where ``$\sim$'' denotes the equivalence for $T$ in the neighbourhood of $+ \infty$. By perturbative analysis we can write for $T$ in the neighbourhood of $+\infty$,
\begin{eqnarray}
\phi_{aT}^{(1)} & = &\phi_a - \frac{\ihbar}{T} \sum_{b \not=a} \frac{\langle \phi_b^*|\dot P_s|\phi_a \rangle}{\lambda_a - \lambda_b} \phi_b + \mathcal O(\frac{1}{T^2}) \\
\phi_{aT}^{*(1)} & = &\phi_a^* - \frac{\ihbar}{T} \sum_{b \not=a} \frac{\langle \phi_b|\dot P^\dagger_s|\phi_a^* \rangle}{\overline{\lambda_a} - \overline{\lambda_b}} \phi_b^* + \mathcal O(\frac{1}{T^2})
\end{eqnarray}
We have then
\begin{equation}
\langle \phi_{aT}^{*(1)}|\dot \phi_{aT}^{(1)} \rangle = \langle \phi_a^*|\dot \phi_a \rangle + \mathcal O(\frac{1}{T})
\end{equation}
\begin{equation}
\frac{\langle \phi_a^{(1)}|\dot \phi_a^{(1)} \rangle}{\langle \phi_a^{(1)}|\phi_a^{(1)} \rangle} = \frac{\langle \phi_a|\dot \phi_a \rangle}{\langle \phi_a|\phi_a \rangle} + \mathcal O(\frac{1}{T})
\end{equation}
\begin{equation}
\langle \phi_{aT}^{*(1)} |H|\phi_{aT}^{(1)} \rangle = \lambda_a + \mathcal O(\frac{1}{T^2})
\end{equation}
\begin{eqnarray}
\frac{\langle \phi_{aT}^{(1)}|H|\phi_{aT}^{(1)}\rangle}{\langle \phi_{aT}^{(1)}|\phi_{aT}^{(1)} \rangle} & = & \lambda_a +  \frac{\ihbar}{T} \sum_{b \not=a} \frac{\langle \phi_a|\phi_b\rangle \langle \phi_b^*|\dot P_s|\phi_a \rangle}{\langle \phi_a|\phi_a \rangle} + \mathcal O(\frac{1}{T^2}) \\
& = & \lambda_a + \frac{\ihbar}{T} \frac{\langle \phi_a|(1-P_s)\dot P_s|\phi_a \rangle}{\langle \phi_a|\phi_a \rangle} + \mathcal O(\frac{1}{T^2})
\end{eqnarray}
and since $P^2_s=P_s \Rightarrow \dot P_s P_s + P_s \dot P_s = \dot P_s \Rightarrow P_s \dot P_s P_s = 0$, we have
\begin{equation}
 \frac{\langle \phi_{aT}^{(1)}|H|\phi_{aT}^{(1)}\rangle}{\langle \phi_{aT}^{(1)}|\phi_{aT}^{(1)} \rangle} = \lambda_a + \frac{\ihbar}{T} \frac{\langle \phi_a|\dot P_s|\phi_a \rangle}{\langle \phi_a|\phi_a \rangle} + \mathcal O(\frac{1}{T^2})
\end{equation}
We see then that for the generator of the dynamical phase we have $\lim_{T \to +\infty} \frac{\langle \underline \psi_T|H|\underline \psi_T \rangle}{\langle \underline \psi_T|\underline \psi_T \rangle} = \lambda_a$, whereas
\begin{eqnarray}
\ihbar^{-1}T \frac{\langle \underline \psi_T|H|\underline \psi_T \rangle}{\langle \underline \psi_T|\underline \psi_T \rangle} & \sim & \ihbar^{-1}T \lambda_a - \frac{\langle \phi_a|\dot P_s|\phi_a \rangle}{\langle \phi_a|\phi_a \rangle} \\
& \sim & \ihbar^{-1}T \lambda_{aT}^{eff}
\end{eqnarray}
with $\lambda_{aT}^{eff} = \lambda_a + \frac{\ihbar}{T} \frac{\langle \phi_a|\dot P_s|\phi_a \rangle}{\langle \phi_a|\phi_a \rangle}$. The deviation between the usual dynamical phase and the effective dynamical phase is precisely equal to the deviation between the adiabatic spectral and orthogonal geometric phases (eq. \ref{deviation}). We have then
\begin{equation}
\ihbar^{-1} T \lambda_a + \langle \phi_a^*|\dot \phi_a \rangle = \ihbar^{-1} T \lambda_{aT}^{eff} + \frac{\langle \phi_a|\dot \phi_a \rangle}{\langle \phi_a|\phi_a \rangle}
\end{equation}
This solves the problem of inconsistency. The adiabatic geometric phases are not equal, but their deviation is compensated by a deviation between the dynamical phases if  $\lambda_{aT}^{eff}$ generates the dynamical phase associated with the orthogonal geometric phase. $\lambda_{aT}^{eff}$ is indeed the correct equivalent of the dynamical phase associated with the nonadiabatic geometric phase. It is interesting to note that the geometric contribution to the dissipation $\left|e^{-\int_0^s \langle \phi_a^*|\dot P_o|\phi_a \rangle ds} \right|^2$ can be then interpreted as a contribution of the dynamical phase.

\section{Conclusion}
Even if the adiabatic spectral geometric phase seems to be more natural with respect to the adiabatic theorem, it is important to note that it is not the adiabatic limit of the nonadiabatic geometric phase; consequently it contributes to the dissipation process. In contrast, the adiabatic orthogonal geometric phase does not contribute to the dissipation process and is thus a good equivalent to a phase for the nonselfadjoint dynamics. This can be very important for experimental measurements of the geometric phase in dissipative quantum dynamics. It not evident that we can have access to a measurement of the adiabatic spectral geometric phase because of its involvement in the quantum flow loss. The adiabatic orthogonal geometric phase could be more pertinent for an experimental measurement.\\

Finally, we can remark that we can also introduce ``non-natural'' geometric phases. Let $\chi(s)$ be a state such that $\langle \chi|\phi_a \rangle \not=0$. $P_\chi = \frac{|\phi_a \rangle \langle \chi|}{\langle \chi|\phi_a \rangle}$ constitutes a projector onto $\Lin(\phi_a)$. A geometric phase generated by $\frac{\langle \chi|\dot \phi_a \rangle}{\langle \chi|\phi_a \rangle}$ is associated with this projection, and we have $\langle \phi_a^*|\dot \phi_a \rangle - \frac{\langle \chi|\dot \phi_a \rangle}{\langle \chi|\phi_a \rangle} = \langle \phi^*_a | \dot P_\chi | \phi_a\rangle = - \frac{\langle \chi| \dot P_s | \phi_a \rangle}{\langle \chi|\phi_a \rangle}$ which is small at the adiabatic limit. If the orthogonal geometric phase has a physical interpretation (it preserves the norm evolution), an interpretation of the non-natural geometric phases is not directly evident (note that the non-natural geometric phase are forbidden in the selfadjoint case, because of the requirement of norm preservation).  Nevertheless, we can say that the geometric phase can be transformed by an arbitrary new kind of gauge change of the form $ \langle \phi^*_a | \dot P_\chi | \phi_a\rangle$ (the usual gauge change being of the form $g^{-1} \dot g$ where $g$ is a non-zero complex number). This remark is particularly interesting since a previous work \cite{viennot} has shown that for some geometric phases associated with a resonance, the geometric structure describing the geometric phase is not a principal bundle (where the only gauge changes are $g^{-1} \dot g$) but a gerbe (which includes also another kind of gauge changes). In ref. \cite{viennot}, the other kind of gauge change is $\langle \phi_a^*|\Omega^{-1} \dot \Omega | \phi_a \rangle$ where $\Omega$ is a wave operator. We remark that in the present case we have $\Omega = P_s (P_\chi P_s P_\chi)^{-1} = P_\chi$ and $\Omega^{-1} = P_\chi P_s = P_s$ (where $(P_\chi P_s P_\chi)^{-1}$ is the inverse in the space spaned by $P_\chi$ and $\Omega^{-1}$ is the weak left inverse of $\Omega$ i.e. $\Omega^{-1} \Omega = P_\chi$). We have then $\langle \phi_a^*|\Omega^{-1} \dot \Omega |\phi_a \rangle = \langle \phi_a^*|\dot P_\chi |\phi_a \rangle$.


\section*{References}
\begin{thebibliography}{0}
\bibitem{messiah} Messiah A 1959 {\it Quantum mechanics} (Paris: Dunod).
\bibitem{marzlin} Marzlin K P and Sanders B C 2004 {\it Phys. Rev. Lett.} {\bf 93}, 160408.
\bibitem{wu} Wu Z and Yang H 2005 {\it Phys. Rev. A} {\bf 72}, 012114.
\bibitem{sarandy} Sarandy M S, Wu L A and Lidar D A 2004 {\it Quantum Information Processing} {\bf 3}, 331.
\bibitem{pati} Pati A K and Rajagopal A K 2005 {\it preprint} arXiv: quant-ph/0405129.
\bibitem{amin} Amin M H S 2009 {\it Phys. Rev. Lett.} {\bf 102}, 220401.
\bibitem{nenciu} Nenciu G and Rasche G 1992 {\it J. Phys. A: Math. Gen.} {\bf 25}, 5741.
\bibitem{abou} Abou Salem W K and Fr\"ohlich J 2007 {\it Commun. Math. Phys.} {\bf 273}, 651.
\bibitem{abou2} Abou Salem W 2007 {\it Ann. H. Poincar\'e} {\bf 8}, 569596.
\bibitem{joye} Joye A 2007 {\it Commun. Math. Phys.} {\bf 275}, 139.
\bibitem{avron} Avron J E, Fraas M, Graf G M and Grech P 2011 {\it Comm. Math. Phys.} {\bf 305}, 633.
\bibitem{avron2} Avron J E, Fraas M, Graf G M and Kenneth O 2011 {\it New J. Phys.} {\bf 13}, 053042.
\bibitem{aharonov} Aharonov Y and Anandan J 1987 {\it Phys. Rev. Lett.} {\bf 58}, 1593.
\bibitem{page} Page D N 1987 {\it Phys. Rev. A} {\bf 36}, 3479.
\bibitem{montragon} Mondrag\'on A and Hern\'andez A 1996 {\it J. Phys. A: Math. Gen.} {\bf 29}, 2567.
\bibitem{mostafazadeh} Mostafazadeh A 1999 {\it Phys. Lett. A} {\bf 264}, 11.
\bibitem{mailybaev} Mailybaev A A, Kirillov O N and Seyranian A P 2005 {\it Phys. Rev. A} {\bf 72}, 014104.
\bibitem{mehri} Mehri-Dehnavi H and Mostafazadeh A 2008 {\it J. Math. Phys.} {\bf 49}, 082105.
\bibitem{viennot} Viennot D 2009 {\it J. Math. Phys.} {\bf 50}, 052101.
\end{thebibliography}
\end{document}